# Real Spectra in PT Symmetry Hamiltonians Using Tridiagonal Representation Approach

T. J. Taiwo

*Department of Physics, United Arab Emirate University, Al ain, UAE*

*Email: Tunde.Joseph@uaeu.ac.ae, Tundetaiwo31@gmail.com*

We consider the solution of the $PT$ symmetry Hamiltonian $H = p^2 + m^2x^2 - (ix)^{2N}$ using the technique of tridiagonal representation approach. This methodology provides more accurate results and proper depictions of the Hamiltonian energy levels and wavefunctions. It is well known that $PT$ symmetry condition of a Hamiltonian ensure that its spectra are real and positive even if the Hamiltonian is non Hermitian. Here, we introduce the method of TRA to get the eigenvalues and wave function of this Hamiltonian for integer values of N and show an approximation solution for the non-integer values of $N$. Due to the nature of the Hamiltonian, the TRA was applied in a semi – analytic manner in this paper.

*In celebration of my 40th Birthday*

## 1. INTRODUCTION

Consider the PT symmetry Hamiltonians

$$H = p^2 + m^2x^2 - (ix)^{2N} \tag{1}$$

where $N \geq 1$ is real. The Hamiltonians is expected to be Hermitian when $N$ is an integer and non Hermitian when $N$ is non- integer. In reference [1], similar Hamiltonians

$$H = p^2 + m^2x^2 - (ix)^{N} \tag{2}$$

was considered there but the solution provided there was the case $m = 0$ and a short comment on the case $m \neq 0$. Equation (2) can also be Hermitian or non Hermitian depending on the value of $N$. Since every Hermitian Hamiltonian (with the usual definitions) is PT symmetric, which ensure that the eigenvalue are real but the converse is not true as can be seen from above from equation (1) and (2).

A quick check for equation (1) lies as follows. For instance, if we write

$$(ix)^{2N} = Exp\left[2N\ln(ix)\right] \tag{3}$$

Using the usual (principal branch) logarithms, we have

$$\ln(ix) = \ln(x) + i\arg(ix) \tag{4}$$

For positive $x$, $\arg(ix) = \pi/2$, so that

$$(ix)^{2N} = x^{2N}Exp[i\pi N] \tag{5}$$

Therefore the potential term becomes

$$-(ix)^{2N} = -x^{2N}Exp[i\pi N] \tag{6}$$

for hermiticity, we must have

$$-\left(x^{2N}Exp[i\pi N]\right)^* = -x^{2N}Exp[-i\pi N] \tag{7}$$

Equating this two equation (6) and (7) give

$$Exp[2i\pi N] = 1 \tag{8}$$

This condition can only be true if $N$ is an Integer. For non-integer $N$, this is not real, so the potential is complex and the Hamiltonian is non Hermitian. However, the $PT$ symmetry nature of the Hamiltonian ensure that the eigenvalues of the Hamiltonian are real for integer values of $N$.

## 2. METHOD

Now we intend to solve equation (1) using Tridiagonal representation approach where we use the oscillator basis element. In brevity, the Tridiagonal Representation Approach can be described as follow, given a differential equation

$$D\psi(x) = \left[p(x)\frac{d^2}{dx^2} + q(x)\frac{d}{dx} + g(x)\right]\psi(x) = 0 \tag{9}$$

The solution is written as the following bounded convergent series

$$\psi(x) = \sum_{n=0}^{\infty} f_n\phi_n(y) \tag{10}$$

where $y = y(x)$ is an independent variable transformation, $\{\phi_n(y)\}$ is a complete set of square integrable functions, and $\{f_n\}$ are the expansion coefficients. The bases are required to carry a tridiagonal matrix representation for the differential operator $D$. That is, the action of $D$ on the basis element should read

$$D\phi_n(y)=\omega(y)\left[a_n\phi_n(y)+b_{n-1}\phi_{n-1}(y)+c_n\phi_{n+1}(y)\right] \quad (11)$$

where $\omega(y)$ is a node-less entire function and $\{a_n,b_n,c_n\}$ are constant coefficients. Hence, the differential equation, $D\psi(x)=0$, becomes a three term recursion relation for the expansion coefficients $\{f_n\}$ that reads

$$a_nF_n+c_{n-1}F_{n-1}+b_nF_{n+1}=0 \quad (12)$$

where we write $f_n=f_0F_n$ making $F_0=1$. Consequently, the solution to the differential equation (9) reduces to an algebraic solution of the discrete relation (12). The set $\{f_n\}$ contains all the properties of the solution $\psi(x)$ as mentioned earlier. There are two solution scenarios for (12)

a. $b_n,c_n>0$, for all $n=0,1,2,...,$

b. $b_n,c_n>0$ only for $n=0,1,2,..N'$

For the first case, $\{F_n\}_{n=0}^{\infty}$ becomes an infinite set of orthogonal polynomial with the following generalized orthogonality (continuous and discrete)

$$\int_{z_-}^{z_+}\rho(z)F_n(z)F_m(z)dz+\sum_{k=0}^{K}\zeta(z_k)F_n(z_k)F_m(z_k)=\xi_n\delta_{n,m} \quad (13)$$

where $z$ is some proper function of the differential equation parameters with $a_n=u_n-zh_n$ such that $\{u_n,h_n,c_n,b_n\}$ are independent of $z$. Moreover, $\xi_n>0$, $K$ is finite or infinite, and $\rho(z)$ is proportional to $\left[f_0(z)\right]^2$. Obviously, the basis set $\{\phi_n(y)\}$ must also be infinite in this case. Sometimes, the solution is either purely continuous (no sum in the orthogonality) or purely discrete (no integral in the orthogonality). For the second case, $\{F_n\}_{n=0}^{N'}$ becomes a finite set of orthogonal polynomial and the basis set $\{\phi_n(y)\}$ must also be finite. The Tridiagonal Representation Approach has two modes of applications, and we have extensively used them it in solving various second order linear differential equations [2-4].

In atomic units, $\hbar=m=1$, with the Hamiltonian in equation (1), the Schrodinger equation becomes

$$\left[-\frac{1}{2}\frac{d^2}{dx^2}+x^2-a\left(x^2\right)^N\right]\psi=E\psi \quad (14)$$

where $a=(i)^{2N}$. This definition of $a$ ensure that for integer $N$, the Hamiltonians become Hermitian and vice versa. Using the Oscillator basis element

$$\phi_n(y)=A_ny^{\alpha}e^{-\beta y}L_n^{v}(y) \quad (15)$$

where $L_n^v(y)$ is generalized Laguerre polynomial, $v>-1$, $\alpha$ and $\beta$ are the basis parameters, and $A_n$ is the normalization constant.

With coordinate transformation $y=\lambda^2x^2$, equation (14) becomes equation (16) as shown on the next page, where $b=\frac{(i)^{2N}}{2\lambda^{2N+2}}$, $\lambda$ is dimensionless parameter. In equation (14) we have used the differential equation of the Laguerre polynomial (Appendix A). Choosing the basis parameters as $\alpha=\frac{v}{2}+\frac{1}{4}$, $\beta=\frac{1}{2}$, and $v=\frac{1}{2}$. We have equation (17) where $d=\left(\frac{1}{4}-\frac{1}{2\lambda^4}\right)$ and subsequently equation (18). Since

$$y^N=N!\sum_{k=0}^{N}\frac{(-1)^k}{(N-k)!}L_k^{1/2}(y) \quad (19)$$

Then equation (18) can therefore be written as equation (20), where we have used the orthogonality property of Laguerre polynomial. The normalization constant is $A_n=\sqrt{\frac{2\lambda\Gamma[n+1]}{\Gamma[n+3/2]}}$. Then, equation (20) will finally be in the form of equation (21). The left hand side of equation (21) gives the Hamiltonian matrice (full matrice) from which one can easily calculate the eigenvalue. With the parameters, $\lambda=1$ and $N=1$, gives a symmetric Hermitian matrix ( 5 by 5 ) in 3 decimal places

$$\begin{pmatrix} 3.096 & -1.879 & 0.335 & 0.060 & 0.024 \\ -1.879 & 8.776 & -4.367 & 0.677 & 0.105 \\ 0.335 & -4.367 & 15.479 & -7.371 & 1.079 \\ 0.060 & 0.676 & -7.371 & 22.975 & -10.805 \\ 0.024 & 0.105 & 1.079 & -10.805 & 31.143 \end{pmatrix} \quad (22)$$

And when $N=1.1$, gives a symmetric non Hermitian matrix

$$\begin{pmatrix} 3.08+0.27i & -1.92-0.42i & 0.36+0.12i & 0.06+0.02i & 0.03+0.01i \\ -1.92-0.42i & 8.86+1.17i & -4.49-1.09i & 0.71+0.23i & 0.11+0.04i \\ 0.36+0.11i & -4.49-1.09i & 15.72+2.42i & -7.60-1.94i & 1.14+0.37i \\ 0.06+0.02i & 0.71+0.23i & -7.60-1.94i & 23.40+3.95i & -11.16-2.94i \\ 0.03+0.01i & 0.11+0.04i & 1.12+0.37i & -11.16-2.94i & 31.79+5.70i \end{pmatrix} \quad (23)$$

$$-\frac{1}{2\lambda^2}(H-E)\left|\phi_n(y)\right\rangle = A_n y^{\alpha} e^{-\beta y}\left\{\begin{array}{l}\left(\begin{array}{l}\left[-(v+1-y)+(2\alpha-2\beta y)+\frac{1}{2}\right]\frac{n}{y}-\frac{\alpha}{y}-n+y\left(\frac{\alpha}{y}-\beta\right)^2\\ +\frac{1}{2}\left(\frac{\alpha}{y}-\beta\right)+by^N-\frac{y}{2\lambda^4}+\frac{E}{2\lambda^2}\end{array}\right)L_n^v(y)\\ -\left[-(v+1-y)+(2\alpha-2\beta y)+\frac{1}{2}\right]\frac{n+v}{y}L_{n-1}^v(y)\end{array}\right\} \quad (16)$$

$$-\frac{1}{2\lambda^2}(H-E)\left|\phi_n(y)\right\rangle = A_n y^{1/2} e^{-1/2y}\left\{\left(-\frac{3}{4}-n+dy+by^N+\frac{E}{2\lambda^2}\right)L_n^v(y)\right\} \quad (17)$$

$$-\frac{1}{2\lambda^2}\left\langle\phi_m\right|(H-E)\left|\phi_n(y)\right\rangle = \frac{A_n A_m}{2\lambda}\int_0^\infty y^{1/2}e^{-y}\left(-\frac{3}{4}-n+dy+by^N+\frac{E}{2\lambda^2}\right)L_n^{1/2}(y)L_m^{1/2}(y)dy \quad (18)$$

$$\begin{array}{l}-\frac{1}{2\lambda^2}\left\langle\phi_m\right|(H-E)\left|\phi_n(y)\right\rangle = \frac{bA_n A_m}{2\lambda}\int_0^\infty y^{1/2}e^{-y}\left(N!\sum_{k=0}^{N}\frac{(-1)^k}{(N-k)!}L_k^{1/2}(y)\right)L_n^{1/2}(y)L_m^{1/2}(y)dy+\left(\frac{E}{2\lambda^2}-\frac{3}{4}-n\right)\delta_{nm}\\ +d\left[(2n+v+1)\delta_{nm}-\sqrt{n(n+v)}\delta_{n,m+1}-\sqrt{(n+1)(n+v+1)}\delta_{n,m-1}\right]\end{array} \quad (20)$$

$$\begin{array}{l}-\frac{(i)^{2N}}{\lambda^{2N}}\frac{A_n A_m}{2\lambda}\int_0^\infty y^{1/2}e^{-y}\left\{N!\sum_{k=0}^{N}\frac{(-1)^k}{(N-k)!}L_k^{1/2}(y)\right\}L_n^{1/2}(y)L_m^{1/2}(y)dy+\\ 2\lambda^2\left\{\left(\frac{3}{4}+n-d\left(2n+\frac{3}{2}\right)\right)\delta_{nm}+d\left[\sqrt{n(n+1/2)}\delta_{n,m+1}+\sqrt{(n+1)(n+3/2)}\delta_{n,m-1}\right]\right\}=E\delta_{nm}\end{array} \quad (21)$$

***Table 1**:* Matrix size 5 by 5 of equation (21) where $\lambda = 2.9$ *, energy levels and eigenvalues.*

| *N* | ***n*** | **Our Case** | **Reference 1** |
|---|---|---|---|
| **4** | 0 | 1.4868 | 1.4771 |
| | 1 | 6.6219 | 6.0333 |
| | 2 | 16.6386 | 11.8023 |
| | 3 | 32.1948 | 18.4590 |

With a proper and accurate variation of our dimensionless parameter $\lambda$, similar eigenvalues to reference 1 can be obtained as shown in table 1. For example, note that when $N = 2$ in our case corresponds, to $N = 4$ in reference [1] but in our case $m \neq 0$. Also equation (21) show similar phenomenon to figure 1 of reference [1]; as $N < 1$ the eigenvalues are not real. As $N \geq 1$, the eigenvalues changes between completely by been totally real or a mixture of finite real and finite complex eigenvalues.

The matrices given by (21) are symmetric and hence can be converted to a tridiagonal matrix by using the ***Householder Transformation method or Lanczos Algorithm.*** Since Hermitian and non Hermitian symmetric matrices have same eigenvalues as like their tridiagonal forms, but it is quite difficult to convert non Hermitian symmetric matrices to their tridiagonal forms, we will consider two cases - ***Case 1*** will deal with integer value of $N$ which will give a symmetric Hermitian matrix and we get the eigenvalues and the correspomding wavefunction for some energy levels. ***Case 2***, we work with non-integer value of $N$ get the eigenvalues, but with some assumption, get the wavefunction for some energy levels.

### *c. RESULTS*

***Case 1: Integer value N = 2 (Matrix size 10 by 10)***

Since the expansion coefficients in the wavefunction are expected to be real orthogonal polynomials, we convert equation (21), a full symmetric Hermitian matrix to a tridiagonal matrix. Therefore, we can write equation (21) as

$$EP_n(E) = a_n P_n(E) + b_{n-1}P_{n-1}(E) + b_n P_{n+1}(E) \quad (24)$$

where $f_n(E) = f_0(E)P_n(E)$ such that $f_0(E) = 1$. It is quite difficult to know the exact orthogonal polynomial that satisfies equation (24). However, numerically it is possible. From equation (21), we have

$$\begin{pmatrix} a_0 - E & b_0 & & & & & \\ b_0 & a_1 - E & b_1 & & & & \\ & b_1 & a_2 - E & b_2 & & & \\ & & b_2 & \ddots & \ddots & & \\ & & & \ddots & \ddots & \ddots & \\ & & & & \ddots & \ddots & \ddots \\ & & & & & \ddots & \ddots & b_{N1-1} \\ & & & & & & b_{N1-1} & a_{N1-1} - E \end{pmatrix} \tag{25}$$

In equation (25), $N1$ is the size of the matrix and $E$ are the eigenvalues of the Hamiltonians. Initial values of orthogonal polynomial is $P_0(E) = 1$, then we can have $P_1(E) = \frac{E - a_0}{b_0} P_0$. With these two values, generally, we can have

$$P_{n+1}(E) = \frac{(E - a_n)P_n(E) - b_{n-1}P_{n-1}(E)}{b_n} \tag{26}$$

for each eigenvalue $E$, we can calculate $P_{N'}(E)$ where $n = 1, 2, ..., N'$ is large enough.

The wavefunctions for each energy level is periodic with their respective zero crossing, and have different spatial distributions. There is no degeneracy as can be seen below in the table 2. Below are the graphs for the case $N = 2$ shown on page 5.

***Table 2**: Matrix size 10 by 10 of Equation (21), few energy levels and eigenvalues in descending order and $\lambda = 3.0$.*

| $N$ | **n** | $E$ |
|---|---|---|
| **2** | **10** | **137.71213** |
| | **9** | **101.06423** |
| | **8** | **74.149449** |
| | **7** | **52.958604** |
| | **6** | **35.996420** |
| | **5** | **22.527162** |

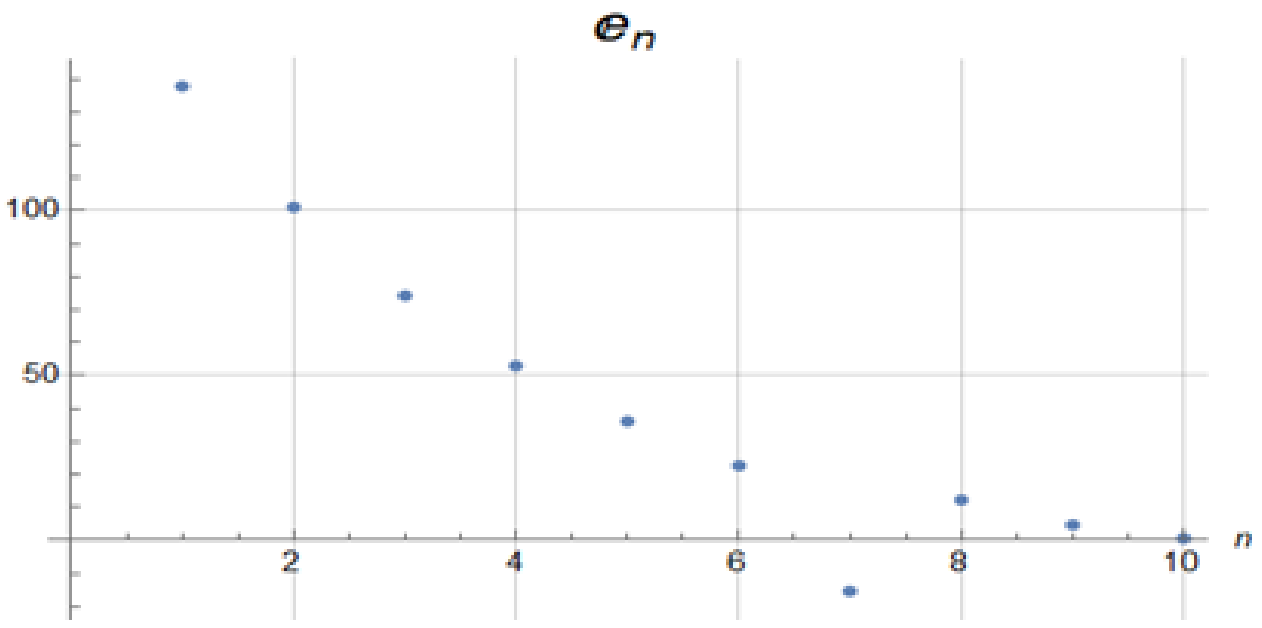

***Figure 1**.* Case 1 - The diagram show how the eigenvalues are distributed among the energy level. A transition is observed as mentioned in reference [1] figure 3 for the massive case.

***Case 2: Integer value N = 0.5 (Matrix size 5 by 5)***

Here, we considered the case when $N = 0.5$. Of course equation (21) will produce complex matrix with no real eigenvalues. In the case, we take the matrix size as 5 by 5. Table 3 shows the eigenvalues to 2 decimal places. There is no restriction to the matrix size we just decided to make it an order of 5 by 5.

***Table 3**: Matrix size 5 by 5 of Equation (21), energy levels and eigenvalues in descending order and $\lambda = 2.5$*

| $N$ | **n** | $E$ |
|---|---|---|
| **0.5** | **4** | **42.57-0.29i** |
| | **3** | **25.03-0.40i** |
| | **2** | **13.84-0.44i** |
| | **1** | **6.62-0.48i** |
| | **0** | **2.28-0.42i** |

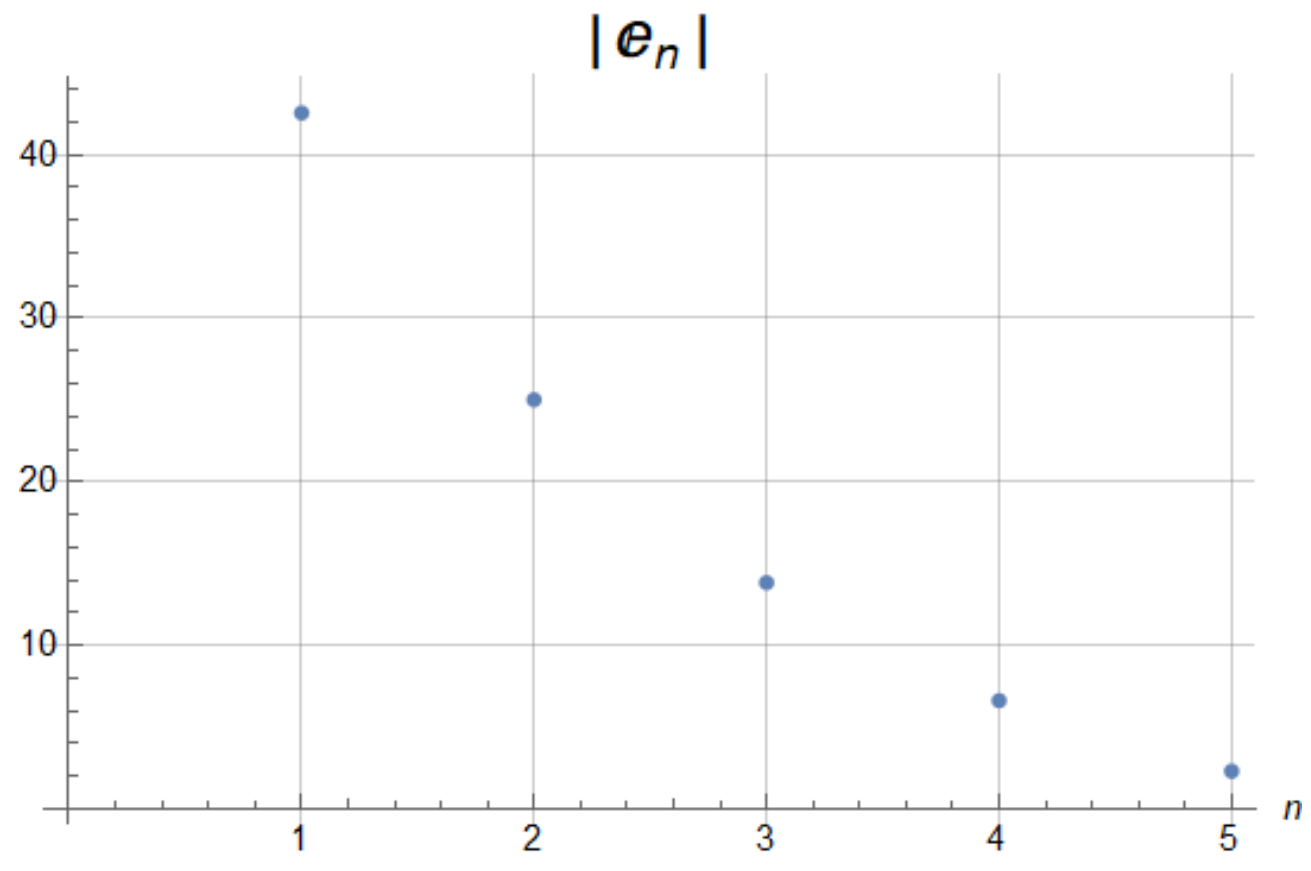

***Figure 3.*** Case 2 - The diagram show how the eigenvalues are distributed among the energy level.

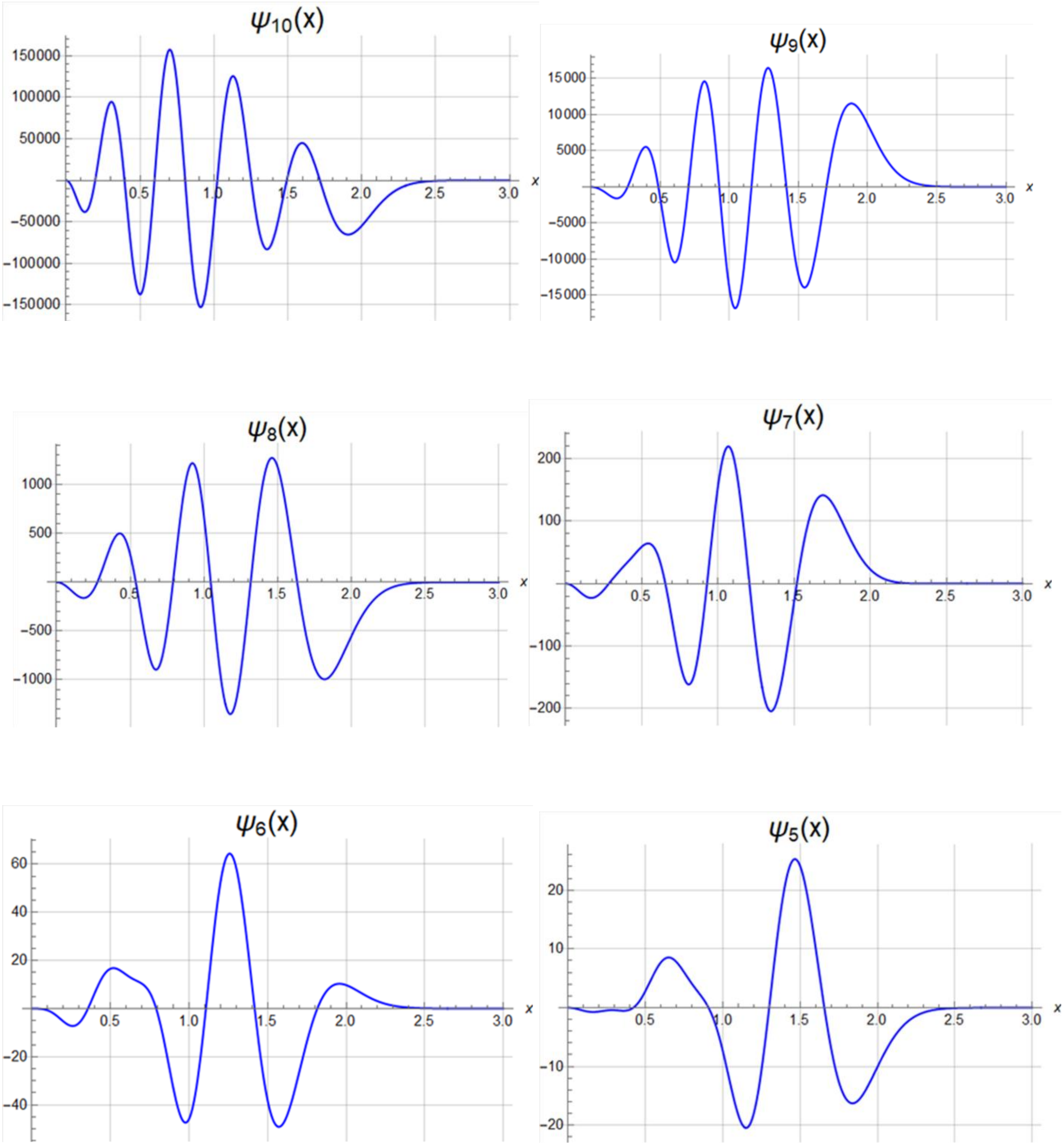

***Figure 2***: These are the graphs of wavefunctions for the first six energy levels in descending order. The matrix size is 10 by 10 with parameter $N = 2$ and $\lambda = 3.0$. The boundary condition that $\psi(x) \to 0$ as $x \to \infty$ is satisfied.

The matrix elements obtained are symmetric non Hermitian matrix. Transforming them into a tridiagonal matrix is difficult but possible. There equation (26) will be a complex orthogonal polynomial like the complex Laguerre polynomial, Complex Hermite Polynomials, Gegenbauer complex Polynomials, Szegő Polynomials, a Bergman Polynomials and so on.

So equation (26) will have a similar form like

$$\tilde{P}_{n+1}(z,\tilde{z}) = \frac{(\tilde{E} - \tilde{a}_n)\tilde{P}_n(z,\tilde{z}) - \tilde{b}_{n-1}P_{n-1}(z,\tilde{z})}{\tilde{b}_n} \qquad (27)$$

with complex values. We don't know the exact complex orthogonal polynomial that will satisfies equation (21) in this case but interestingly these polynomials have an initial values of $\tilde{P}_n(0) = 1$. Therefore we can use the entries of the matrix elements along the diagonal $\tilde{a}_n(\tilde{z})$, the eigenvalues, and with the upper and lower off diagonal elements which might not be exact as $\tilde{b}_n(\tilde{z})$ if the matrix has been transformed to a tridiagonal form. However, we plot the absolute values of the wavefunction as shown above in figure 3.

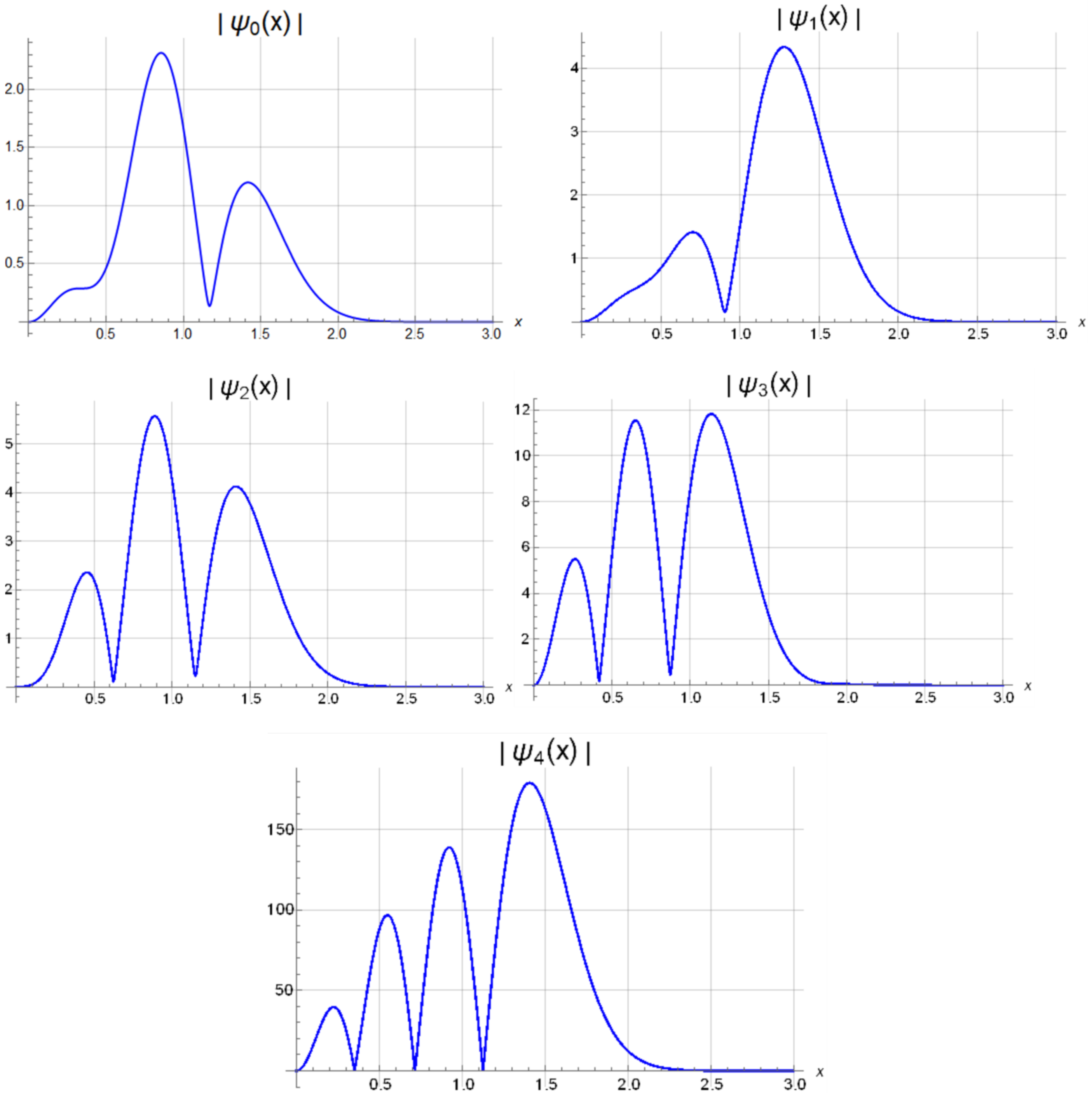

***Figure 3***: These are the graphs of absolute wavefunctions for the first six energy levels in descending order. The matrix size is 5 by 5 with parameter $N = 0.5$ and $\lambda = 2.5$. The boundary condition that $\psi(x) \to 0$ as $x \to \infty$ is satisfied

## 4. *CONCLUSION*

We introduced the method of Tridiagonal representation approach to the study of non Hermitian *PT* symmetry Hamiltonians. Lot of works had been done on the PT symmetry and Quantum physics, for example by CM Bender etal [8]. Since the TRA is an algebraic method of solution that has been used extensively in quantum physics, it is needful to see how TRA can contribute to the study of PT symmetry and its advantages. Here, we studied a *PT* symmetry Hamiltonian defined in equation (1) and successfully find the respective energy level and wavefunction. All relevant physical phenomenons are depicted in comparism with similar work about the PT symmetry Hamiltonian. We plan to follow up by considering other PT symmetry Hamiltonians and see how the TRA can be of a better approach to other methods.

In conclusion, the Tridiagonal Representation Approach (TRA) is an algebraic technique for resolving second-order linear ordinary differential equations. The method's algebraic advantage is strengthened by the ability to analyze special functions and orthogonal polynomials. It is preferred on the computational side since it relies on strong numerical methods that handle Gauss quadrature and tridiagonal matrices. The method uses a whole set of square integrable functions (basis) to express the solution of a second order linear differential equation (such as the Schrödinger equation) as a convergent series, either infinite or finite. This set is selected so that the differential operator's matrix representation is tridiagonal.

All physical properties of the solution (e.g., energy spectrum, scattering phase shift, density of states, etc.) are obtained from the properties of these polynomials (e.g., their weight function, generating function, zeros, asymptotics, etc.).

## 5. *ACKNOWLEDGEMENT*

The author appreciates the financial and technical support Saudi Centre for theoretical physics and United Arab Emirates University department of Physics.

## APPENDIX A

Laguerre Basis: This basis element is given as

$$\left|\phi_n(r)\right\rangle = A_n y^{\alpha} e^{-\beta y} L_n^{v}(y) \tag{A1}$$

where $\alpha$ and $v$ are real parameters with $v > -1$ and $\alpha \geq 0$ to ensure convergence of the Laguerre polynomial $L_n^v(y)$ and compatibility with boundary conditions when the new variable $y$ span semi-infinite interval $[0,\infty]$. Also, in choosing $A_n$ insight is required based on the derivatives of the configuration space transformation coordinate. The Laguerre polynomial $L_n^v(y)$ satisfies the following properties:

$$yL_n^v(y) = (2n+v+1)L_n^v(y) - (n+v)L_{n-1}^v(y) - (n+1)L_{n+1}^v(y) \tag{A2}$$

$$L_n^v(y) = \frac{\Gamma(n+v+1)}{\Gamma(n+1)\Gamma(v+1)}\,{}_1F_1(-n; v+1, y) \tag{A3}$$

$$\left[y\frac{d^2}{dy^2} + (v+1-y)\frac{d}{dy} + n\right]L_n^v(y) = 0 \tag{A4}$$

$$y\frac{d}{dy}L_n^v(y) = nL_n^v(y) - (n+v)L_{n-1}^v(y) \tag{A5}$$

$$\int_0^\infty y^v e^{-y} L_n^v(y) L_m^v(y)\,dy = \frac{\Gamma(n+v+1)}{\Gamma(n+1)}\delta_{nm} \tag{A6}$$